\documentclass[a4paper]{article}
\usepackage{graphicx}

\begin{document}

\begin{center}
{\Large \bf Toward UrQMD Model Description of $pp$ and $p{\rm C}$ Interactions at High Energies }
\end{center}

\begin{center}
{V. Uzhinsky}
\end{center}

\begin{center}
{Laboratory of Information Technologies, JINR, Dubna, Russia}
\end{center}

\begin{center}
\begin{minipage}{12cm}
It is found that UrQMD model version 3.3 does not describe NA61/SHINE Collaboration data
on $\pi$-meson production in $pp$ interactions at energies 20 -- 80 GeV. At the same time,
it describes quite well the NA49 Collaboration data on the meson production in $pp$ and $p{\rm C}$
interactions at 158 GeV. The Collaborations do not consider feedback of $\eta$-meson decays.
All versions of the UrQMD model assume that $\eta$-mesons are "stable". An inclusion of the decays
into calculations leads to 2--3 \% increase of the meson production which is not enough for
description of the data. Possible ways of the model improvements are considered.\\
    Conclusions of the paper are: accounting of $\eta$-meson decays is not essential for
a description of experimental data; a new tuning of the UrQMD model parameters is needed
for a successful description of $pp$ and $p{\rm C}$ interactions at high energies; inclusion of
the low mass diffraction dissociation in the UrQMD model would be desirable.
\end{minipage}
\end{center}

\section{$pp$ interactions}
The NA61/SHINE Collaboration is going to obtain a high precision data on $\pi^-$-meson inclusive
cross sections in $pp$ interactions at $P_{lab}=$ 20, 31, 40, 80 and 158 GeV/c. Some preliminary
results are published in the paper \cite{NA61}. The NA49 Collaboration has published the high
precision data on $\pi^\pm$, $K^\pm$, proton and antiproton spectra in $pp$ \cite{NA49pp2pi,NA49pp2p,NA49pp2K}
and $p{\rm C}$ \cite{NA49pC} interactions at 158 GeV/c. The NA61/SHINE Collaboration has the analogous
data for $p{\rm C}$ interactions at 31 GeV/c \cite{NA61pC}.
There were some attempts to describe the last data in the modern Monte Carlo models -- FLUKA, VENUS,
FTF and UrQMD \cite{NA61pC,UzhiImp,UzhiTune}. As it was shown in the paper \cite{UzhiImp}
the $p{\rm C}$ data at 31 GeV/c can be described in the UrQMD model after a small improvement of the model
code. An analysis of the above mentioned data can point on other improvements of the model.

The UrQMD model \cite{UrQMD} is a tool for analysis of nucleus-nucleus interactions
at high energies. It is implemented as a Monte Carlo code \cite{UrQMDpage} which is widely used at
design of future experiments at FAIR \cite{FAIR} and NICA \cite{NICA} facilities. Its validation for
hadron-nucleon interactions presented at HEPWEB server \cite{HEPWEB} is not complete, and
can be improved now.
\begin{figure}[cbth]
\includegraphics[width=80mm,height=45mm,clip]{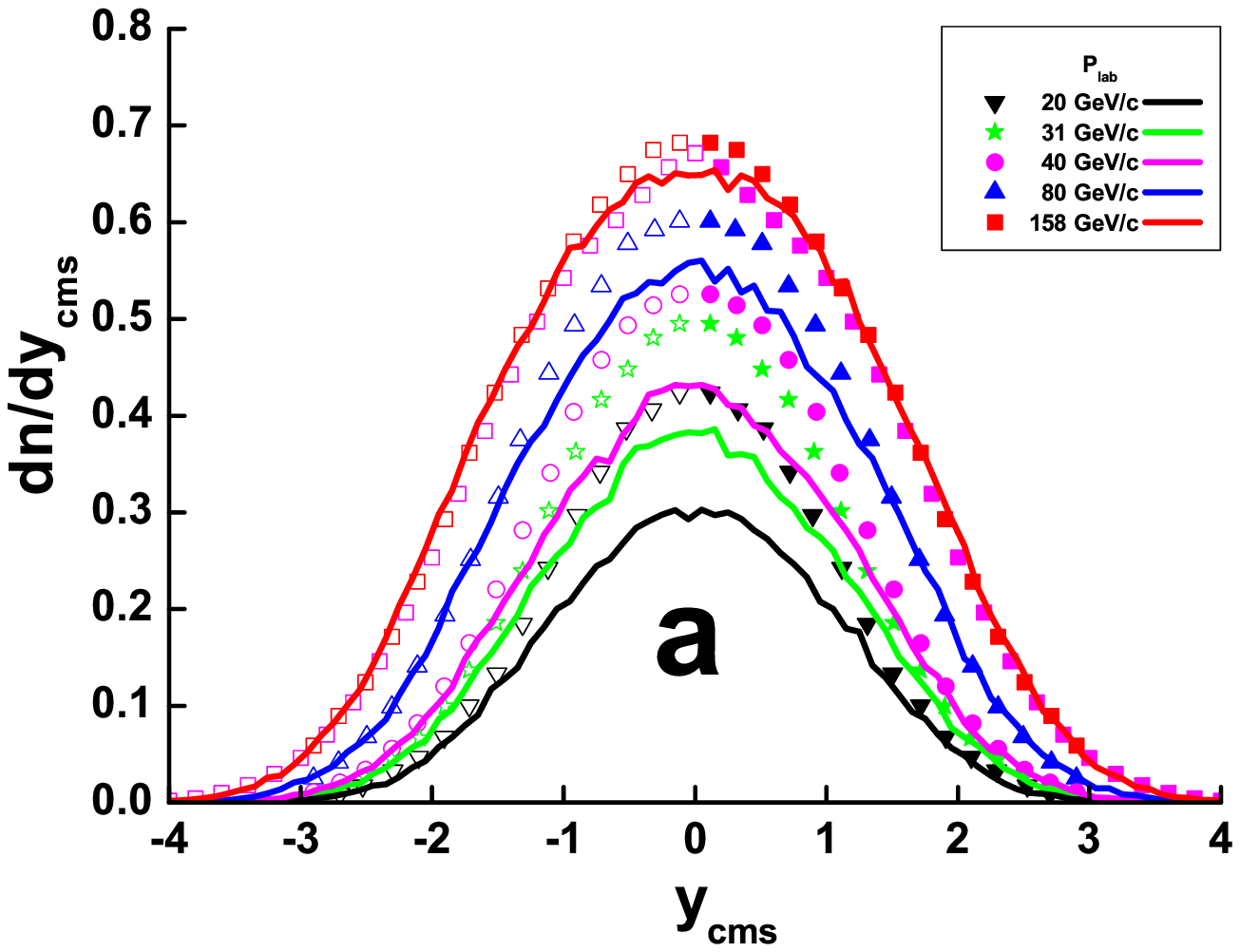}\includegraphics[width=80mm,height=45mm,clip]{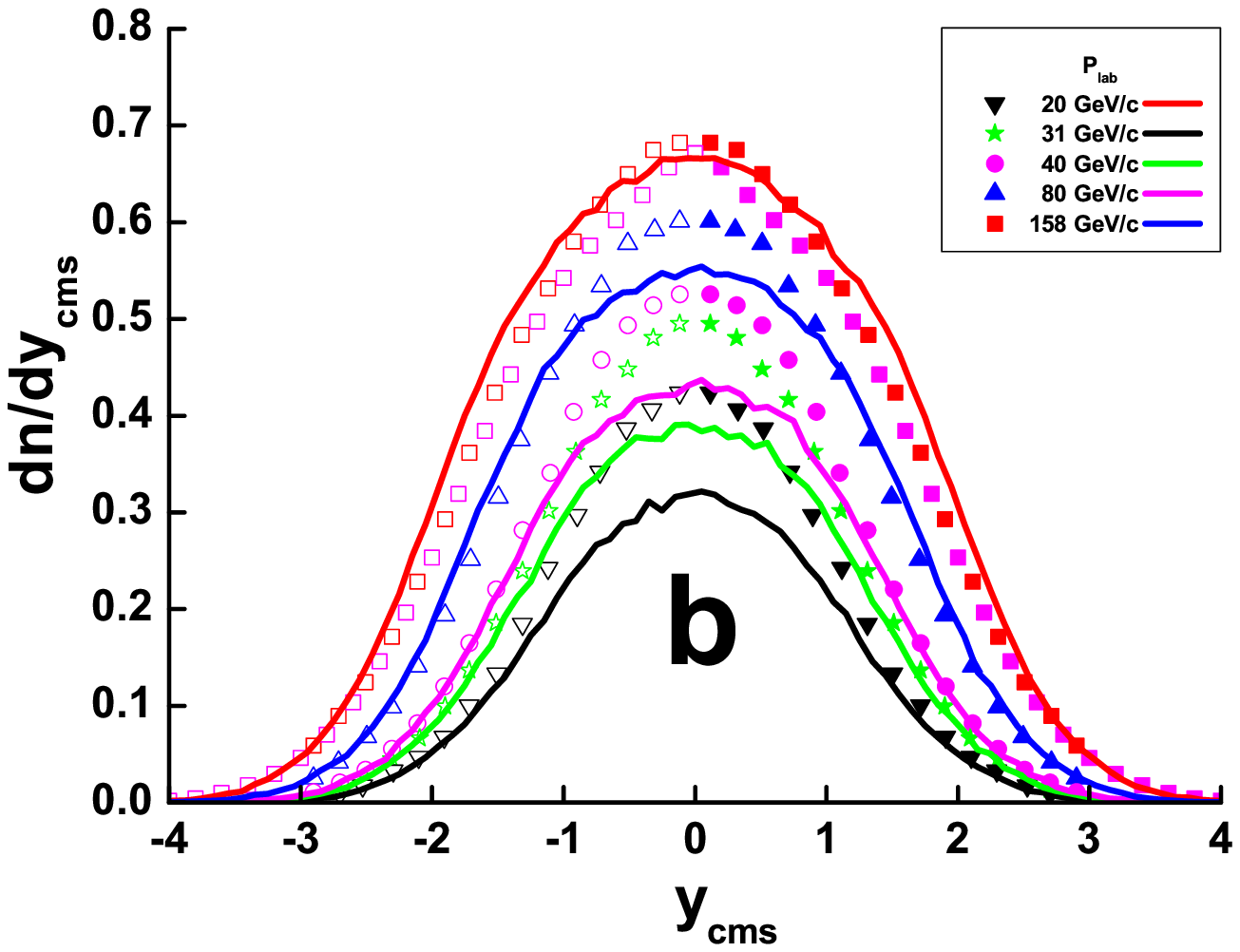}
\caption{Rapidity distributions of $\pi^-$-mesons in $pp$ interactions. Closed points are preliminary
NA61/SHINE data \protect{\cite{NA61}}, the open points are the data reflected at mid-rapidity.
Magenta squares are the NA49 data at 158 GeV/c \protect{\cite{NA49pp2pi}}.
Lines are the UrQMD 3.3 model calculations: a) calculations with default values of the model
parameters; b) calculations with $\eta$-meson decays.}
\label{NA61pp}
\end{figure}

As it is shown in Fig.~1a the model version 3.3 with default values of the parameters cannot describe the
NA61/SHINE Collaboration data on $pp$ interactions. Only at $P_{lab}=$158 GeV/c there is an agreement
between the model calculations and the NA49 data.

At the calculations, $\eta$-mesons were considered as "stable"ones. At the same time the Collaborations
did not take into account feedbacks of the $\eta$-meson decays. An inclusion of the decays in the calculations
leads to few percent increase of the spectra in central regions (see Fig.~1b) which is not sufficient for
a good description of the data.

The same conclusion about the role of the $\eta$-meson decays can be done at an application
of the model to the NA49 data \cite{NA49pp2pi,NA49pp2K} (see Fig.~2).
\begin{figure}[cbth]
\includegraphics[width=160mm,height=100mm,clip]{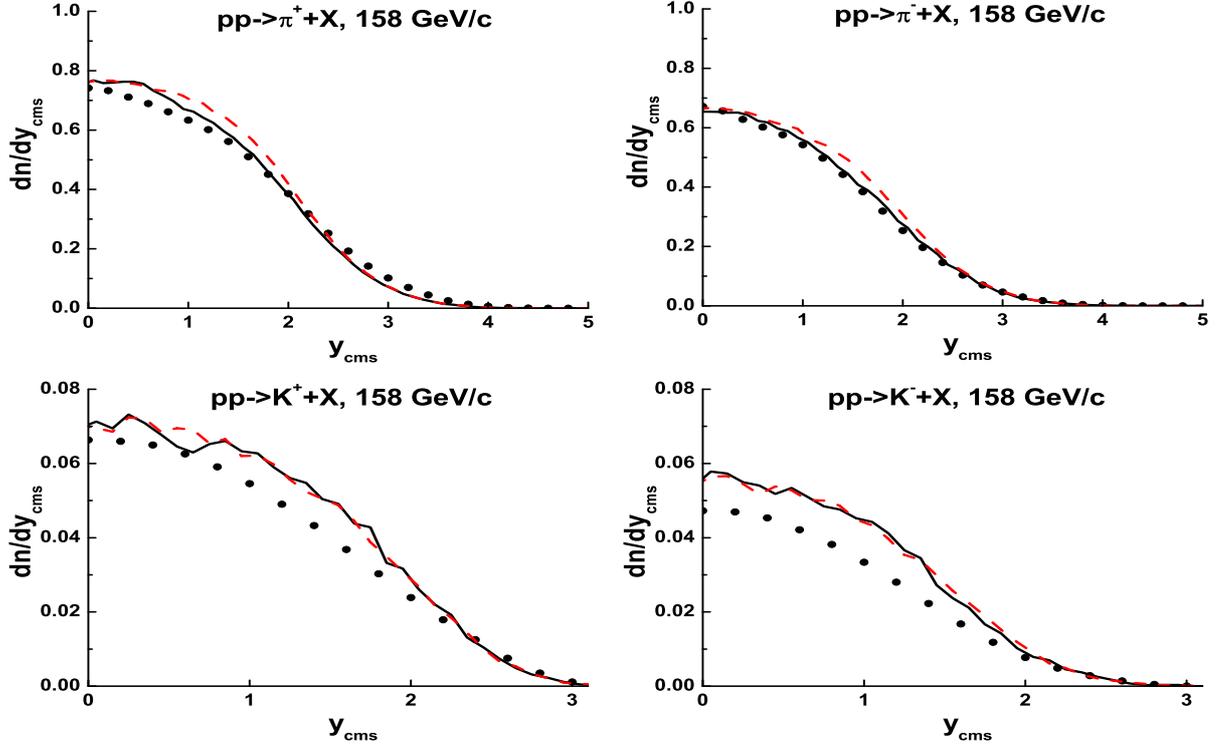}
\caption{Rapidity distributions of mesons in $pp$ interactions at $P_{lab}=$ 158 GeV/c.
Points are experimental data \protect{\cite{NA49pp2pi,NA49pp2K}}. Lines are UrQMD 3.3 model calculations:
solid lines are the calculations with default values of the parameters, dashed lines are the
calculations with accounting of $\eta$-meson decays.}
\label{Xtotel}
\end{figure}

The following changes were introduced in the UrQMD model code (file urqmd.f) for a consideration of
the decays:
\begin{verbatim}
c
c  Main program
c
      lbm(8,102)=1               ! Setting decays of eta-meson              ! Uzhi
      branmes(0,102)=0.39d0      ! Setting decay to gg                39%   ! Uzhi
      branmes(8,102)=0.61d0      ! Setting decay to 3 Pi              61%   ! Uzhi

      mc=0
      mp=0
      noc=0
c
c loop over all events
c
\end{verbatim}

The following lines were added and closed below in the file:
\begin{verbatim}
c is particle unstable
*            if(dectime(i).lt.1.d30) then                      ! Uzhi
            if((dectime(i).lt.1.d30).or.(ityp(i).eq.102)) then !Forcing eta-meson decay
                                                               ! Uzhi
 41            continue
               isstable = .false.
               do 44 stidx=1,nstable
                  if (ityp(i).eq.stabvec(stidx)) then
c                     write (6,*) 'no decay of particle ',ityp(i)
                     isstable = .true.
                  endif
 44            enddo
               if (.not.isstable) then
c     perform decay
                  call scatter(i,0,0.d0,fmass(i),xdummy)
c     backtracing if decay-product is unstable itself
*                 if(dectime(i).lt.1.d30) goto 41                  ! Uzhi
             if((dectime(i).lt.1.d30).or.(ityp(i).eq.102)) goto 41 ! Forcing eta-meson decay
                                                                   ! Uzhi
               endif
            endif
c     check next particle
            if(i.lt.npart) goto 40
         endif ! final decay
         CTOption(10)=CTOsave
c final output
\end{verbatim}

As known, the UrQMD model uses the Fritiof model \cite{Fritiof} for simulations of hadronic interactions
at high energies. It also simulates various binary reactions like that $NN\rightarrow N\Delta$,
$NN\rightarrow NN^*$, $NN\rightarrow N\Delta^*$, $NN\rightarrow \Delta\Delta$, $NN\rightarrow \Delta N^*$
and $NN\rightarrow \Delta\Delta^*$, $\Delta N^*$. They are very important at low energies. At the considered
energies the Fritiof processes are mainly responsible for particle production in the central region, at
$y_{cms}\sim 0$. The above mentioned binary processes give contributions in the projectile and target
fragmentation regions. Thus, a simple way to improve the situation is a decreasing of the binary reaction
cross sections. The decreasing in 3 times gives results presented in Fig.~3.
\begin{figure}[cbth]
\includegraphics[width=160mm,height=80mm,clip]{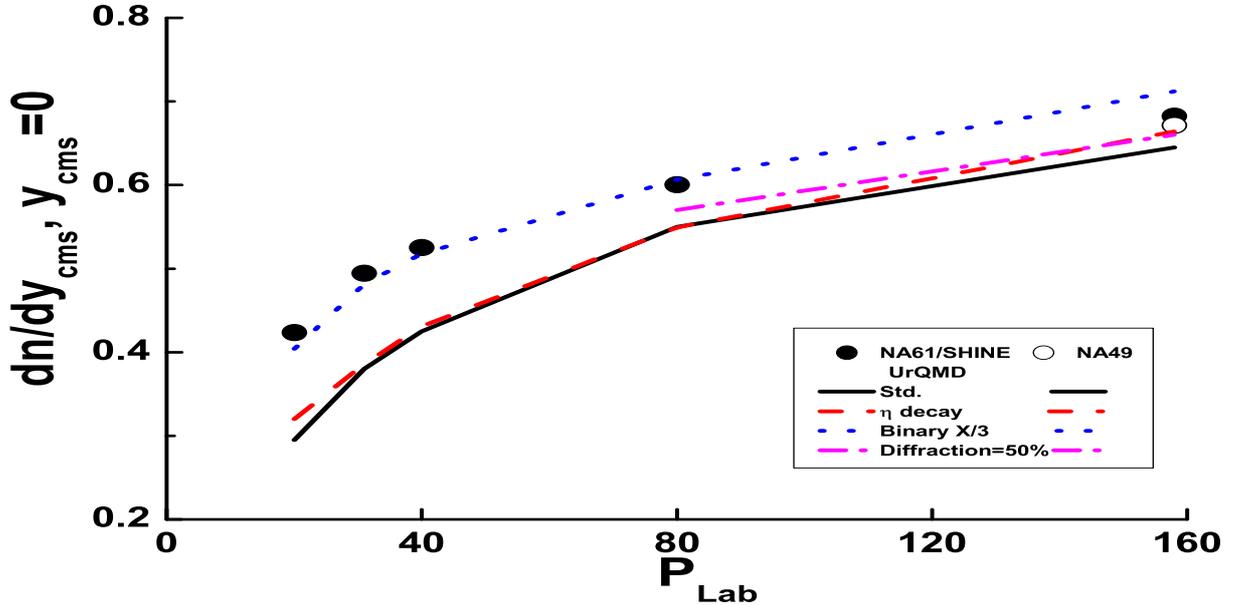}
\caption{Energy dependence of meson inclusive cross sections in $pp$ interactions in the central region.
Points are experimental data \protect{\cite{NA61,NA49pp2pi}}. Lines are the UrQMD 3.3 model calculations:
solid line is a calculation with default parameters; dashed line is a calculation with accounting of
the $\eta$-meson decays; dotted line is a calculation with decreased binary reaction cross sections;
dot-dashed line is a calculation with 50 \% probability of the single diffraction dissociation.}
\label{dNdY0}
\end{figure}

As seen, the decreasing allows to describe meson production at 20, 31, 40 and 80 GeV/c. At larger energies
it leads to an overestimation of the inclusive cross sections. Because the cross sections of the binary
reactions are small at high energies, the main contribution to the spectra is connected with the Fritiof
processes. The processes are single diffraction dissociation and "non-diffractive" interaction (more
exactly, two-vertex diffraction dissociation).

A probability of the single diffraction is a decreasing function of an energy in the Fritiof model
which contradicts with known experimental data. Thus, the spectra at mid-rapidity are growing faster
than it is needed. This can be erased fixing the probability at 50 \%, for example. As seen in Fig.~3
the assumption
allows to describe the data at 158 GeV/c. The calculations underestimate the data at 80 GeV/c.
Summing up, one can conclude that accurate parameterizations of the binary reaction cross sections and
the single diffraction cross sections are needed for a correct description of experimental data on
$pp$ interactions.

The following changes were introduced in the file scatter.f to implement the decreasing of the binary
reaction cross sections:
\begin{verbatim}
               call crossx(iline,sqrts,ityp1,iso31,fmass(ind1),
     &                     ityp2,iso32,fmass(ind2),sigma(ii-2))
       if((ityp1.le.100).and.(ityp2.le.100).and.(ii.gt.4).and.(ii.le.10) ! Uzhi
     & .and.(sqrts.ge.3.5d0)) sigma(ii-2)=sigma(ii-2)/3.                 ! Uzhi
            else   !  detailed balance:
\end{verbatim}

The implementation of the increased probability of the single diffraction in the file make22.f
is more complicated:
\begin{verbatim}
         if(ranf(0).ge.0.5d0) then                          ! Uzhi
 81      continue
c 100 tries for excitation, otherwise elastic scattering
         ntry=ntry+1
         if(ntry.gt.1000)then
          write(*,*)'make22: too many tries for string exc. ->elastic/
     &           deexcitation'
          write(*,*)' i1, i2, m1, m2, e: ',i1,i2,m1,m2,e
         i1=i1old
         i2=i2old
          goto 13
         endif

c string-excitation:
c get string masses ms1,ms2 and the leading quarks
         call STREXCT(IFdiq1,IFqrk1,ib1,M1m,
     &           ifdiq2,ifqrk2,ib2,M2m,E,
     &           iexopt,
     &           ba1,ms1,ba2,ms2,
     &           ifdiq3,ifqrk3,ifdiq4,ifqrk4)
c the boost parameters are now fixed for the masses ms1, ms2. If the
c masses will be changed, set the parameter fboost to "false":
      fboost=.true.

      if((ms1.le.msmin1).or.(ms2.le.msmin2)) goto 81        ! Uzhi
c accept deexcitation of one of the hadrons:
*      if(ms1.le.msmin1.and.ms2.ge.msmin2)then
*        ms1=massit(i1)
*        fboost=.false.
*      else if(ms2.le.msmin2.and.ms1.ge.msmin1)then
*        ms2=massit(i2)
*        fboost=.false.
c don't accept elastic-like (both masses too low):
*      else if(ms1.lt.msmin1.and.ms2.lt.msmin2)then
*        goto 81
*      endif

      else   !-----------------------------------------     Uzhi
c in case of deexcitation new masses are necessary
c single diffractive, mass excitation according 1/m
*      if(ms1.le.msmin1)then
*        ms2=fmsr(msmin2,e-ms1)
*        fboost=.false.
*      elseif(ms2.le.msmin2)then
*        ms1=fmsr(msmin1,e-ms2)
*        fboost=.false.
*      endif

         call STREXCT(IFdiq1,IFqrk1,ib1,M1m,               ! Uzhi
     &           ifdiq2,ifqrk2,ib2,M2m,E,                  ! Uzhi
     &           iexopt,                                   ! Uzhi
     &           ba1,ms1,ba2,ms2,                          ! Uzhi
     &           ifdiq3,ifqrk3,ifdiq4,ifqrk4)              ! Uzhi

      if(ranf(0).ge.0.5d0) then    !-------- Targ. diffr.  ! Uzhi
        ms1=massit(i1)             !                       ! Uzhi
        ms2=fmsr(msmin2,e-ms1)     !                       ! Uzhi
        fboost=.false.             !                       ! Uzhi
      else                         !-------- Proj. diffr.  ! Uzhi
        ms2=massit(i2)             !                       ! Uzhi
        ms1=fmsr(msmin1,e-ms2)     !                       ! Uzhi
        fboost=.false.             !                       ! Uzhi
      endif                                                ! Uzhi
      endif !-----------------------------------------     ! Uzhi

c quark-quark scattering -> only elastic !
\end{verbatim}
\begin{figure}[cbth]
\includegraphics[width=160mm,height=55mm,clip]{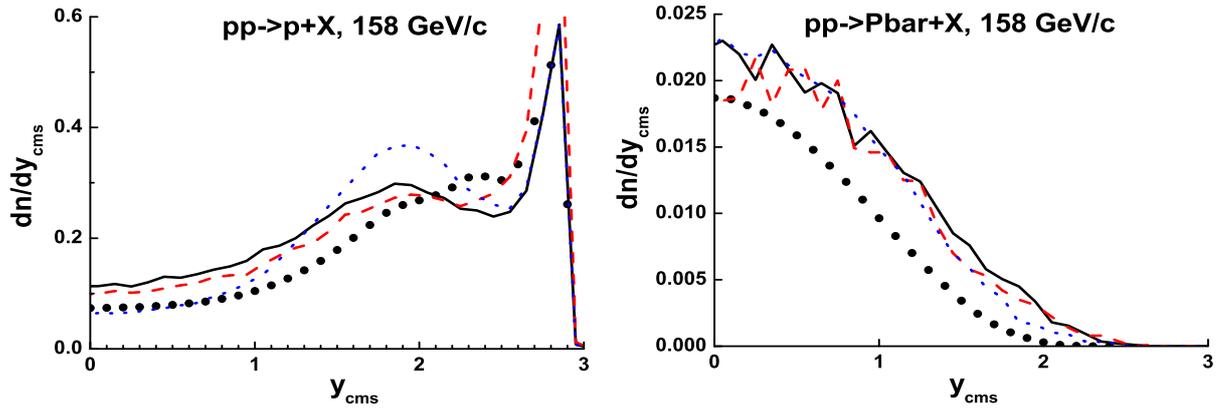}
\caption{Rapidity distributions of protons and antiprotons in $pp$ interactions at 158 GeV/c.
Points are experimental data \protect{\cite{NA49pp2p}}. Lines are the UrQMD 3.3 model calculations:
solid lines are the calculations with default parameters; dashed lines are obtained at the 50 \%
probability of the single diffraction; dotted lines are the calculations at the assumption
that the leading particle effect is valid for all baryons.}
\label{PPy158}
\end{figure}

Very complicated situation takes place with baryon spectra in $pp$ interactions. It is presented in Fig.~4.
As seen, the UrQMD model predicts a dip in the proton rapidity spectrum at $y_{cms}\sim 2.5$ and a maximum
at $y_{cms}\sim 1.75$. A peak at $y_{cms} \sim 2.8$ reflects the single diffraction dissociation.
The experimental spectrum has more simple structure. The main difference between the experimental data
and the model calculation is connected with proton multiplicity in the inelastic $pp$ collisions. As seen,
the model overestimate the multiplicity.

The increasing of the probability of the single diffraction leads to an abnormal increase of
the forward peak and a small decreasing of the proton yield in the central region (see dashed
curves in Fig.~4). The difference between the experimental data and the calculations in the
central region cannot be explained by newly produced protons because the corresponding multiplicity
of newly produced antiprotons is small.

Leading protons and neutrons originated from primary protons are treated in the model differently from other
baryons - $\Delta$, $\Delta^*$, $N^*$ and so on. A special fragmentation function is used for
a production of the leading nucleons. Assuming that all baryons obey the leading particle
effect\footnote{It was implemented in the file string.f opening the commented line
"c~~~~~~~~\& (abs(IDENT(I)).ge.1110))then" and closing the previous line
" \& (abs(IDENT(I)).eq.1120 .or.abs(IDENT(I)).eq.1220) )then".}$^)$
we obtain dotted curves in Fig.~4. As seen, the assumption does not allow to improve
the situation.

In order to find a process what can be responsible for a filling of the dip presented in the proton spectrum
at $y_{cms}\sim 2.5$, let us look at proton spectra in various processes considered by the UrQMD model.
The spectra are shown in Fig.~5.
\begin{figure}[cbth]
\includegraphics[width=160mm,height=100mm,clip]{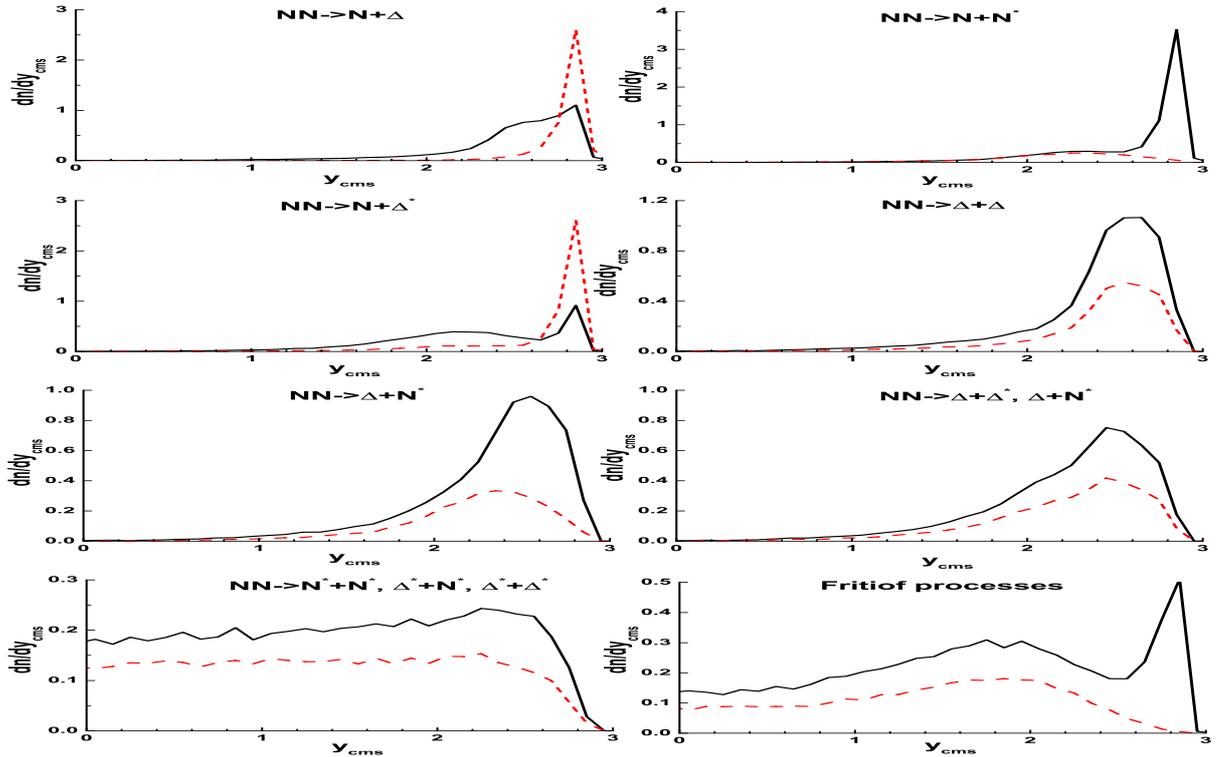}
\caption{Proton and neutron spectra in various processes of the UrQMD model. Solid and dashed lines,
correspondingly.}
\label{Procs}
\end{figure}

As seen, the $NN\rightarrow N\Delta$ process can give an yield in the dip region, but according to
the reggeon phenomenology its cross section must decrease as $1/s$ where $s$ is CMS energy squared.
It is also true for the $NN\rightarrow \Delta\Delta$ process. The processes $NN\rightarrow \Delta N^*$
and $NN\rightarrow \Delta\Delta^*$, $\Delta N^*$ can give the needed contributions. It is expected that
their cross sections decrease as $1/\sqrt{s}$. In the processes, systems with small masses ($M_X <3$--5 GeV)
are created in an association with $\Delta$-resonance. In the reggeon phenomenology a low mass diffraction
dissociation is considered -- $NN\rightarrow N+X$. A mass distribution of system $X$ looks like $1/M_X^2$.
Its cross section is proportional to the elastic cross section, $\sigma_{LMD}\simeq 0.55\ \sigma_{el}$.
Thus, the cross section does not
decrease with energy growth. The diffraction can fill the dip of the proton spectrum. The above
selected processes -- $NN\rightarrow \Delta N^*$ and $NN\rightarrow \Delta\Delta^*$, $\Delta N^*$,
can imitate the diffraction in the UrQMD model. In order to include the diffraction, cross section of
the process $NN\rightarrow \Delta N^*$ was increased in the scatter.f file:
\begin{verbatim}
               call crossx(iline,sqrts,ityp1,iso31,fmass(ind1),
     &                     ityp2,iso32,fmass(ind2),sigma(ii-2))
       if((sqrts.ge.17.2).and.(ii.eq. 8)) sigma(ii-2)=4.d0               ! Uzhi
            else   !  detailed balance:
\end{verbatim}
This improves the proton spectrum in the dip region and $\pi$-meson rapidity spectra (see Fig.~6).
\begin{figure}[cbth]
\includegraphics[width=160mm,height=80mm,clip]{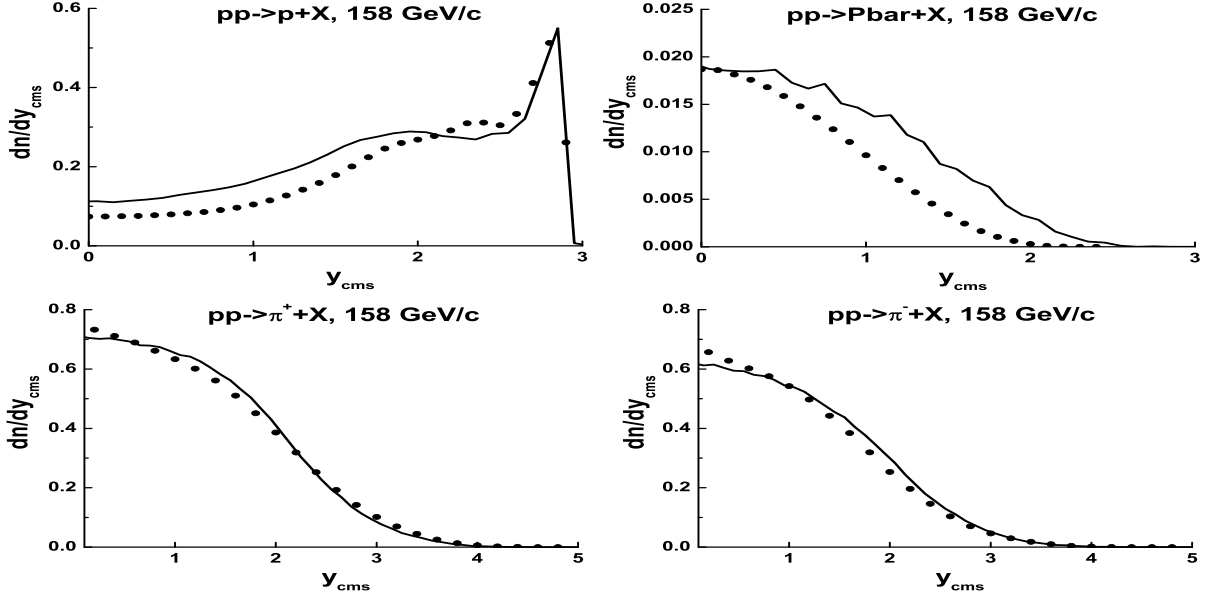}
\caption{Rapidity spectra of particles in $pp$ interactions at 158 GeV/c. Points are the experimental data
\protect{\cite{NA49pp2p}}. Lines are the UrQMD model calculations with imitation of the low mass diffraction.}
\label{LMD}
\end{figure}

Now the main problem is a large protons multiplicity in the final states. The excess of protons in
the model can be explained by a damped multiplicity of strange baryons.
To check the hypothesis we need an exact data on hyperon production in $pp$ interactions.
To a good luck, the NA49 Collaboration in the paper \cite{NA49pp2pi} presented in Fig.~22
$p_T$ integrated density distributions, $dn/dx_F$, of $\Lambda$, $\Sigma^+$ and $\Sigma^-$ hyperons
($x_F=2\ p_z/\sqrt{s}$). The data are shown in Fig.~7 below in a comparison with model calculations.
\begin{figure}[cbth]
\centerline{\includegraphics[width=120mm,height=60mm,clip]{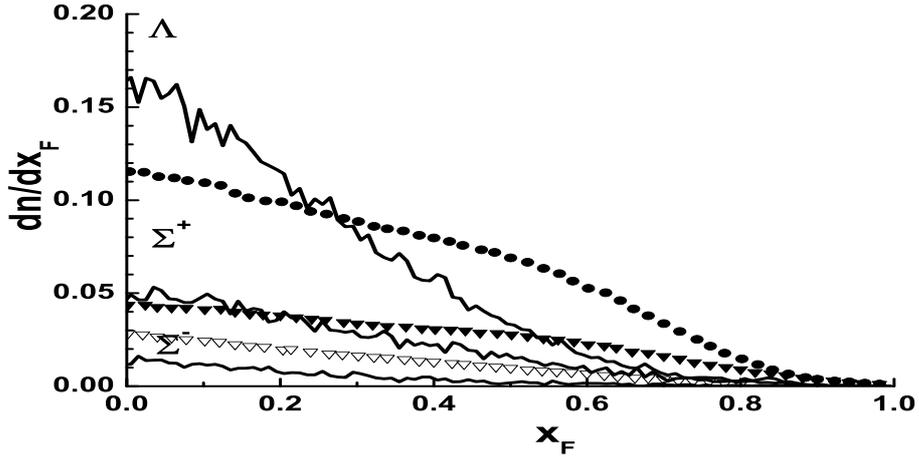}}
\caption{$dn/dx_F$ distributions of $\Lambda$, $\Sigma^+$ and $\Sigma^-$ hyperons in $pp$ interactions
at 158 GeV/c. Points are experimental data \protect{\cite{NA49pp2pi}}. Lines are the UrQMD 3.3 model calculations
with default parameter values.}
\label{Lambda158}
\end{figure}

As seen, the model really underestimates the hyperon production in the most interesting region
$x_F > 0.3$, and overestimates the yield at $x_F\sim 0$. All attempts to improve the description
were not successful. Maybe, the deficit of $\Lambda$ hyperons at $x_F > 0.3$ is caused by the fact
that many $\Lambda$ hyperons are produced in the model -- $\Lambda(1116)$, $\Lambda(1405)$,
$\Lambda(1520)$ and so on. It seems that due to a dominance of heavy $\Lambda$ hyperons and decays
of heavy $\Lambda$ hyperons, $\Lambda(1116)$ has a soft spectrum. Because the simulation algorithm
has a complicated structure it is too difficult to improve the situation.

\section{$p{\rm C}$ interactions}
The UrQMD model with default parameters values and with the accounting of the $\eta$-meson decays
describes rather well the NA49 data on $\pi$-meson production in $p{\rm C}$ interactions at 158 GeV/c
(see solid curves in Fig.~8). At the same time, the model overestimates the proton yield in
the central and target fragmentation regions. The increase of the probability of the single diffraction
to 50 \% for the improvement of the proton spectrum at large rapidities in $pp$ interactions does not
help (see dotted curves in Fig.~8). The peak at $y_{CMS}\sim$ 2.7 is saved, but the proton production
in the target fragmentation region is underestimated. The $\pi$-meson data are underestimated also.
The form of the proton spectrum is not reproduced.

The imitation of the low mass diffraction gives nearly the same results (see dashed lines in Fig.~8).
For implementation of the imitation, the another changes were done in scatter.f in order to introduce
an energy dependence of the process:
\begin{verbatim}
               call crossx(iline,sqrts,ityp1,iso31,fmass(ind1),
     &                     ityp2,iso32,fmass(ind2),sigma(ii-2))
       if((ityp1.le.100).and.{ityp2.le.100).and.(ii.ne.3).and.     ! Uzhi
     &    (ii.le.10).and.(sqrts.ge.3.5d0)) sigma(ii-2)=0.          ! Uzhi
       if((ityp1.le.100).and.{ityp2.le.100).and.                   ! Uzhi
     &    (ii.eq.8).and.(sqrts.ge.3.5d0)) sigma(ii-2)=0.55*sigma(3)! Uzhi

            else   !  detailed balance:
\end{verbatim}
\begin{figure}[cbth]
\includegraphics[width=160mm,height=100mm,clip]{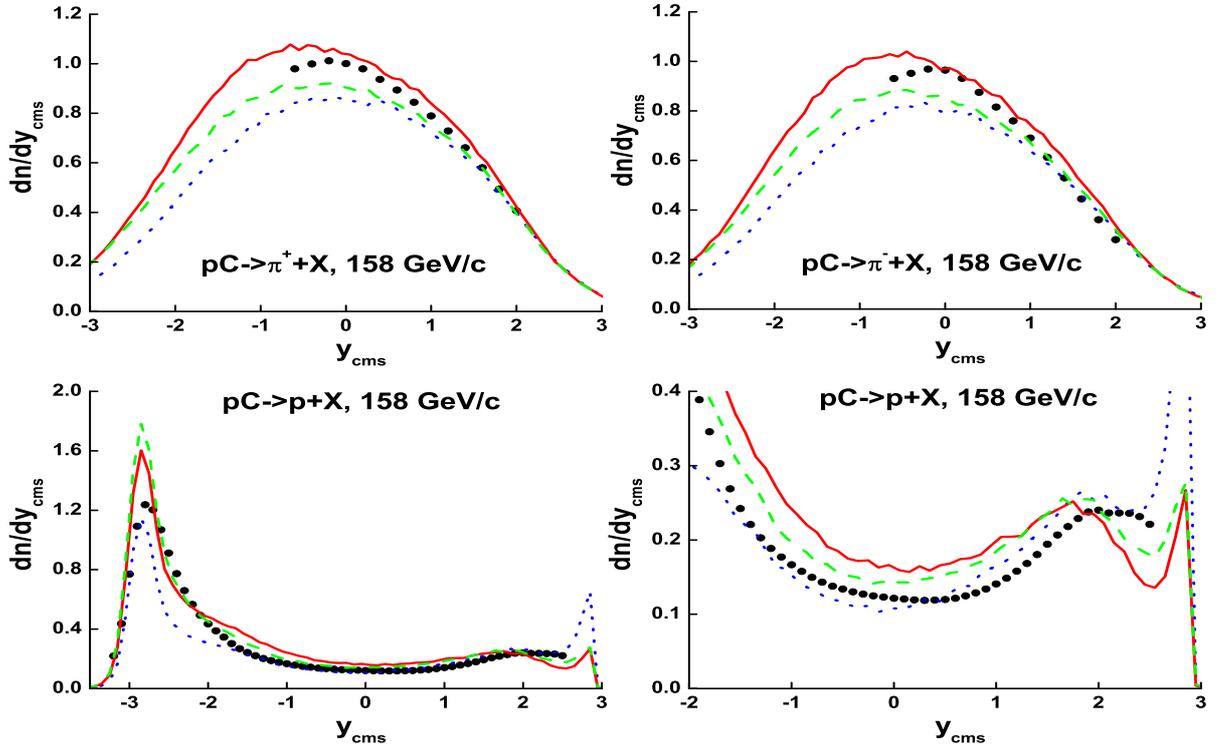}
\caption{Rapidity distributions of $\pi^\pm$-mesons and protons in $p{\rm C}$ interactions
at 158 GeV/c. Points are experimental data \protect{\cite{NA49pC}}. Lines are the UrQMD 3.3 model calculations:
solid lines are the calculations with the accounting of the $\eta$-meson decays; dashed lines are
the calculations with the imitation of the low mass diffraction;
dotted lines are results obtained at the 50 \% probability of the single diffraction.}
\label{pC158}
\end{figure}

Results of simulations of $p{\rm C}$ interactions at 31 GeV are presented in Figs.~9,~10 in a comparison
with the NA61/SHINE data \cite{NA61,NA61pC} by the solid lines. Thin lines are the calculations with
default parameter values and the $\eta$-meson decays. Thick ones are the calculations with the imitation of
the low mass diffraction. As seen, the results are rather close to each other. The calculation results
reasonably agree with the $\pi^+$ data at all angles. The model overestimates the $\pi^-$ data at all angles,
and reproduces qualitative the proton data \cite{NA61}.
\begin{figure}[cbth]
\includegraphics[width=160mm,height=150mm,clip]{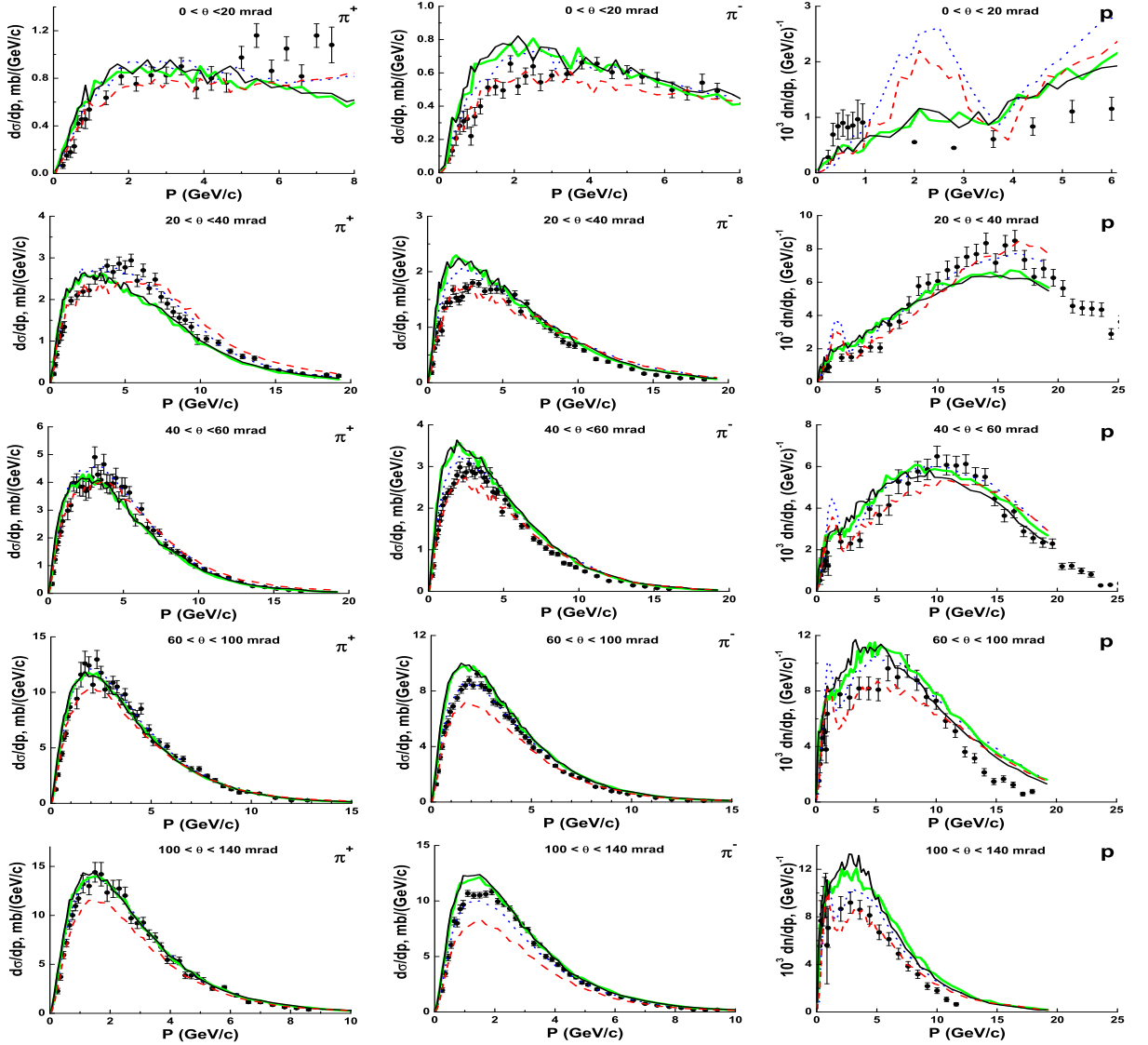}
\caption{Momentum spectra of $\pi^\pm$ and protons in $p{\rm C}$ interactions
at 31 GeV/c. Points are experimental data \protect{\cite{NA61,NA61pC}}. Lines are the UrQMD 3.3 model
calculations with the $\eta$-meson decays: thick solid lines (green) are the calculations with
the imitation of the low mass diffraction; thin solid lines (black) are the calculations with
default parameter values; dashed lines are the calculations for $pp$ interactions
with default values of the model parameters; dotted lines are the calculations for $pp$ interactions
with the imitation of the low mass diffraction.}
\label{pC31a}
\end{figure}

In order to understand the behaviour of the calculations, let us look at results of simulations of
$pp$ interactions re-normalized on an inelastic $p{\rm C}$ cross section (252 mb). The
matter is, an average multiplicity of intra-nuclear collisions in $p{\rm C}$ interaction is close
to one according to the Glauber approximation. Thus, we can expect that properties of $p{\rm C}$
interactions will not be very different from properties of $pp$ interactions. In Figs.~9,~10
dashed lines are the calculations for $pp$ interactions performed with default values of the model
parameters and with accounting of the $\eta$-meson decays. As seen, the calculation underestimates
the $\pi^+$- and $\pi^-$-meson data. Let us remind, that this variant of the calculations gave
the essential underestimation of $\pi^-$-meson production in the central region. This underestimation
reflects in the momentum spectra. The imitation of the low mass diffraction dissociation gives results
close to the data (see dotted curves in Figs.~9,~10). It is very strange, that the two variants of
the calculations give similar results for $p{\rm C}$ interactions.
\begin{figure}[cbth]
\includegraphics[width=160mm,height=150mm,clip]{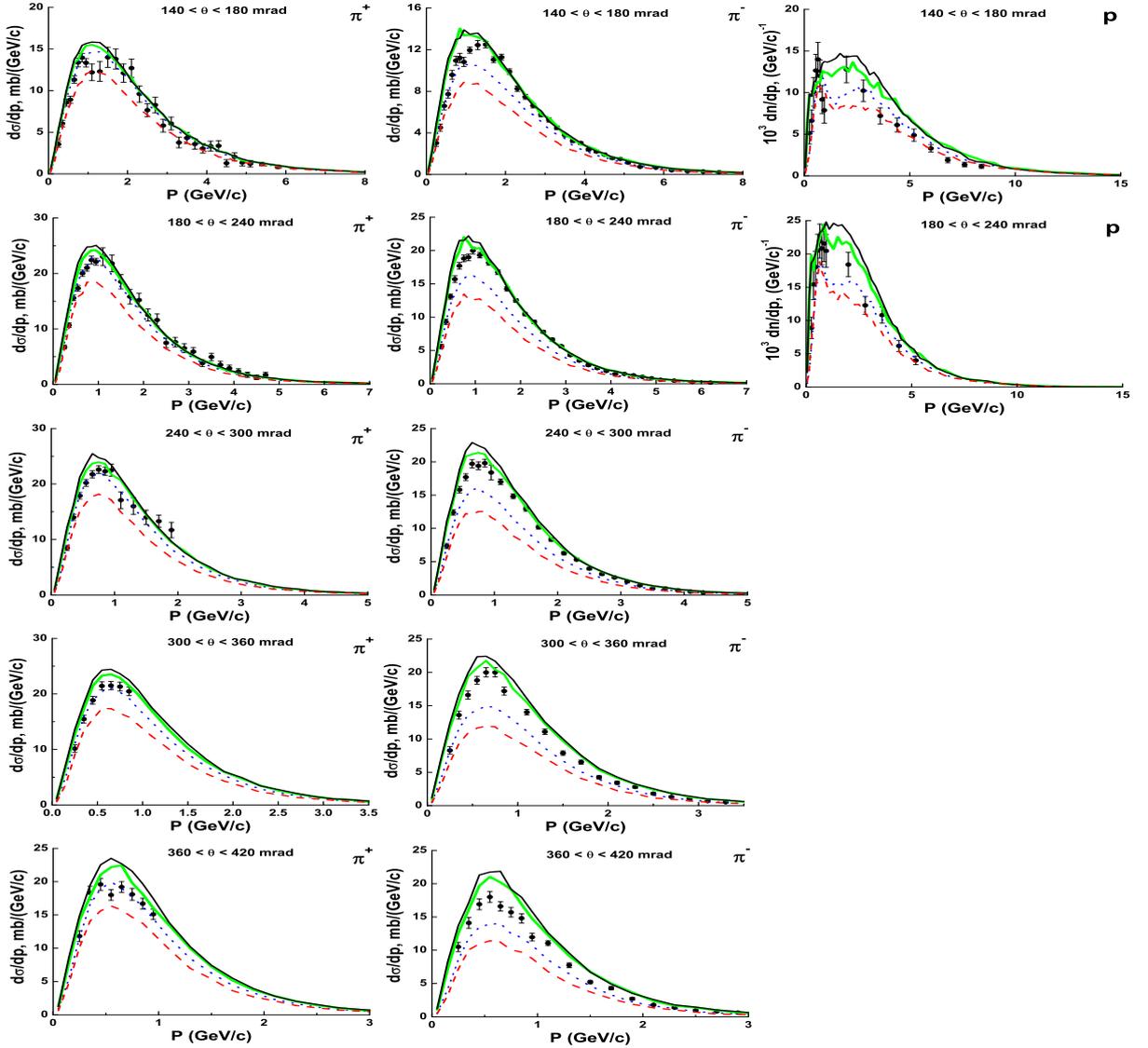}
\caption{Momentum spectra of $\pi^\pm$ and protons in $p{\rm C}$ interactions
at 31 GeV/c. Points are experimental data \protect{\cite{NA61,NA61pC}}. Lines are the UrQMD 3.3 model
calculations with the $\eta$-meson decays: thick solid lines (green) are the calculations with
the imitation of the low mass diffraction; thin solid lines (black) are the calculations with
default parameter values; dashed lines are the calculations for $pp$ interactions
with default values of the model parameters; dotted lines are the calculations for $pp$ interactions
with the imitation of the low mass diffraction.}
\label{pC31b}
\end{figure}

Proton spectra in $pp$ interactions are very interesting. As seen, there is a broad peak in the calculated
proton spectrum at $\theta <$ 20 mrad and $p\sim$ 2 GeV/c. It is directly connected with the binary
reactions or with the low mass diffraction dissociation. It is located in the target fragmentation
region which was not resolved quite well in old bubble chamber experiments (typically, proton identification
was done at $p<$ 1.2 GeV/c). It would be very useful to study the region more carefully. I think,
the NA61/SHINE Collaboration has all opportunities to measure proton spectra in $pp$ interactions in the region.

As seen also in Fig.~9, the peak is practically disappeared in $p{\rm C}$ interactions at
$\theta <$ 20 mrad according to the model calculations. It is explained by the fact, that there is no peak in
the proton spectrum of proton-neutron ($pn$) interactions. $pp$ and $pn$ collisions are presented in equal
amount.

The peak is presented in the model proton spectra of $pp$ interactions at all angles, and is not seen
in the $pn$ calculations. The experimental data
on $p{\rm C}$ interactions show a plateau at $p\simeq$ 2--5 GeV/c and $20 <\theta <$ 140 mrad.
The VENUS model calculations have the plateau very close to the experimental one \cite{NA61} at
$20 <\theta <$ 60 mrad. The FLUKA model gives the peak at small momenta and $20 <\theta <$ 100 mrad
\cite{NA61}. So, there is a problem in other models with accurate simulations of the binary reactions
or the low mass diffraction dissociation in hadron-nucleus interactions. A correct implementation
of the low mass diffraction dissociation at a simulation of hadron-nucleus interactions requires
a lot of efforts, but the theoretical framework (see for example \cite{GIS1,GIS2} and references there)
exists which simplifies the task.

For further analysis it is useful to re-plot the NA61/SHINE data in other form shown in Fig.~11.
There a particle rapidity is used instead of momentum.
\begin{figure}[cbth]
\includegraphics[width=160mm,height=60mm,clip]{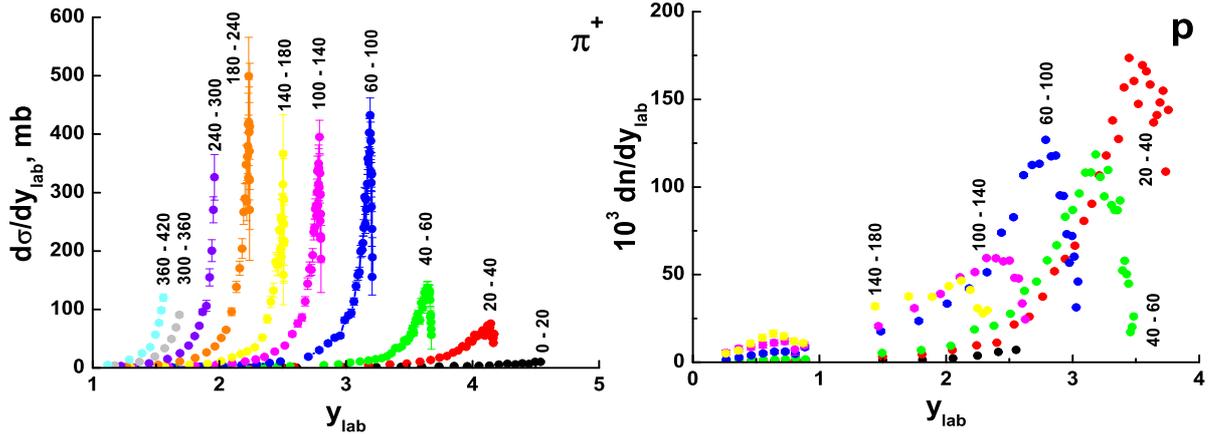}
\caption{Rapidity spectra of $\pi^+$-mesons and protons. Points are the experimental data
\protect{\cite{NA61,NA61pC}}.}
\label{RapPiP}
\end{figure}

Now it is clear, that the $\pi$-meson data at $\theta <$ 60 mrad belong to particles produced
in the projectile fragmentation region. The data at larger angles show the particle production
in the central region.

The proton data at $20 <\theta <$ 60 mrad are relating to the most interesting region where
many proton production mechanisms are mixing in the projectile fragmentation region. The data
at low momenta (less than 5 GeV/c) are connected with proton production in the target fragmentation region.

The projectile and target fragmentation
regions are not determined quite well. In the reggeon phenomenology, their size is about 1 -- 1.5
in rapidity. At projectile momentum 31 GeV/c, the target fragmentation region in $y_{lab}$ extends
from 0 to 1.5, the projectile fragmentation region is from 2.5 to 4. This definition was
used above in the paper.

\section*{Conclusion}
\begin{itemize}
\item Accounting of $\eta$-meson decays is not essential for a description of experimental data.

\item A new tuning of the UrQMD model parameters is needed for a successful description of $pp$ and
$p{\rm C}$ interactions at high energies.

\item Inclusion of the low mass diffraction dissociation in the UrQMD model would be desirable.

\item The experimental data by NA49 and NA61/SHINE Collaborations are very useful for improvement
of the UrQMD model.
\end{itemize}

The author is thankful to A. Galoyan for interest to the work, and to B. Popov for very useful remarks.

\end{document}